# Photoreforming of plastic waste into valuable products and hydrogen using a high-entropy oxynitride with distorted atomic-scale structure

Ho Truong Nam Hai[a,b], Thanh Tam Nguyen[a,c], Maiko Nishibori[d], Tatsumi Ishihara[a,c,e], Kaveh Edalati[a,b,c,]*

[a] WPI, International Institute for Carbon Neutral Energy Research (WPI-I2CNER), Kyushu University, Fukuoka 819-0395, Japan
[b] Department of Automotive Science, Graduate School of Integrated Frontier Sciences, Kyushu University, Fukuoka 819-0395, Japan
[c] Mitsui Chemicals, Inc. - Carbon Neutral Research Center (MCI-CNRC), Kyushu University, Fukuoka, 819-0395, Japan
[d] International Center for Synchrotron Radiation Innovation Smart, Tohoku University, Sendai, Miyagi 980-8577, Japan
[e] Department of Applied Chemistry, Faculty of Engineering, Kyushu University, Fukuoka, 819-0395, Japan

The persistent existence of plastic waste causes serious problems for the environment, directly and indirectly affecting the health of organisms and humans. Photoreforming is a nature-friendly method that only uses solar energy to convert plastic waste into green hydrogen ($H_2$) and valuable organic products. This study shows that a high-entropy oxynitride (HEON) photocatalyst, synthesized by the addition of nitrogen to a Ti-Zr-Hf-Nb-Ta-containing high-entropy oxide (HEO), exhibits a higher potential for the production of $H_2$, formic acid and acetic acid from polyethylene terephthalate (PET) photoreforming compared to the relevant HEO. Examination of X-ray absorption near edge structure (XANES) and extended X-ray absorption fine structure (EXAFS) by synchrotron light shows that, in addition to hybridization of 2p orbitals from oxygen and nitrogen, nitrogen atoms distort the structure and completely change the neighborhood of niobium and titanium (a main contributor to the conduction band), expands the atomic bonds of zirconium and tantalum, contracts the atomic bonds of hafnium and decreases the binding energy of titanium, niobium and tantalum. These electronic structure changes lead to a narrower bandgap and diminished electron-hole recombination, enhancing the photoreforming performance. This study introduces HEONs with distorted atomic-bond structures as efficient low-bandgap and stable catalysts for transforming plastics into high-value organic chemicals and $H_2$ by photocatalysis.

*Corresponding author: Kaveh Edalati (E-mail: kaveh.edalati@kyudai.jp; Tel: +80-92-802-6744)

## 1. Introduction

The consumption of fossil fuel resources and increasing the level of carbon dioxide ($CO_2$) emissions worldwide are leading to energy and environmental crises [1]. Hydrogen ($H_2$) is a clean fuel with high energy density (122 kJ/g) which does not emit $CO_2$ or any other toxic substances, meeting the criteria for a green and sustainable energy carrier [2]. With the sole use of solar energy in photocatalysis, synthesizing $H_2$ from abundant resources of water, biomass and waste by the catalytic process becomes an ideal path for clean fuel production [1,2]. Production of $H_2$ from waste is not only a clean fuel production solution but also a path for managing waste like disposed plastics.

Plastics play a critical role in human life; however, approximately 9-30% of them are recycled [3]. The uncontrolled disposal of plastics into the environment has led to a global plastic pollution problem and the loss of valuable resources [4]. Global efforts toward plastic waste management have recently been enhanced by working on recycling, converting, and replacing plastic with other more environmentally friendly materials. Even so, the most optimistic scenario still shows that plastic emissions will continue to increase at an alarming rate in the coming decades [4–9]. To contribute to reducing plastic pollution, plastic waste collection and recycling programs have been carried out in many countries. Despite the effectiveness of these programs for the collection of large-sized plastic waste, a significant problem is the spread of smaller plastic pieces, known as microplastics and nanoplastics, on the lands, rivers and oceans [7,10]. As for recycling, the process usually changes the mechanical properties through thermal decomposition of polymer chains, resulting in a lower quality of recycled plastics compared to the original one [6]. For example, an assessment showed that only about 7% of used polyethylene terephthalate (PET) plastic bottles can currently be recycled into a new plastic bottle [4]. Plastic products that have been recycled two times or more can hardly be recycled further and should be burned, a process that releases a certain amount of $CO_2$ into the atmosphere [6,11].

Converting plastics into valuable products has become a significant research direction in recent years. Photoreforming is a promising technology that can contribute to the conversion of plastic waste into valuable products on the one hand and create $H_2$ gas as a green fuel on the other hand [12,13]. This technology uses only sunlight and a reusable photocatalyst for the conversion [14]. In the photoreforming process, the catalyst receives photons from a radiation source, leading to the excitation of electrons ($e^-$) to move to the conduction band (CB) from the valence band (VB), and simultaneously producing photogenerated holes ($h^+$) [15]. These charge carriers ($e^-$ and $h^+$) then move to the surface of the catalyst to contribute to $H_2$ production and plastic conversion. The plastic waste is exposed to the photogenerated holes and oxidized to other organic molecules [14]. Meanwhile, excited electrons participate in the reduction process of water to form $H_2$ [4]. The method is clean and quite simple with the capability to be conducted under atmospheric temperature and pressure using solar radiation and a photocatalyst like $TiO_2$ [14], ZnO [16], CdS [4], $Mo_2C$ [17] and $C_3N_4$ [18]. The main challenge for the process is to obtain feasible photocatalysts with a narrow bandgap, wide light absorption, low electron-hole recombination, high activation and a stable structure within many reaction cycles [15,19–21]. Entropy-stabilized materials have a high potential for such a catalytic application.

As a new class of entropy-stabilized materials, high-entropy oxides (HEOs) have received much attention [22–29]. HEOs are composed of at least five principal cations, having a stable crystal structure with entropies greater than $1.5R$, where $R$ is the gas constant [30,31]. The structure of HEOs is usually characterized by severe lattice distortion, the formation of oxygen vacancies, the heterogeneous distribution of valence electrons, slow lattice diffusion and the cocktail effect [30,32]. A large number of HEO catalysts were reported in various catalytic applications such as $Mg_{0.2}Co_{0.2}Ni_{0.2}Cu_{0.2}Zn_{0.2}O$ for lithium batteries [23], $(CeZrHfTiLa)O_x$ for CO oxidation [24], Pt/Ru(NiMgCuZnCo)O for $CO_2$ conversion [25], $TiZrHfNbTaO_{11}$ for $H_2$ generation [26], $La(CrMnFeCo_2Ni)O_3$ [27] and ZnFeNiCuCoRu-O [33] for oxygen evolution, $(FeCoNiCuZn)WO_4$ for plastic degradation [28], and $(Eu_{1-x}Gd_x)_2(Ti_{0.2}Zr_{0.2}Hf_{0.2}Nb_{0.2}Ce_{0.2})_2O_7$ for radioactive waste treatment [29]. Recently, the combination of the concept of HEO and high-entropy nitrides (HENs) in a few publications has introduced a new class of catalysts, namely high-entropy oxynitrides (HEONs) for catalytic [15] or mechanical [34] applications. However, there have been no studies on the effectiveness of HEONs for the photoreforming of plastic waste and comparing their activity with corresponding HEOs by considering their atomic structure differences.

In this study, a Ti-Zr-Hf-Nb-Ta-based HEON is used for the photoreforming process to degrade plastics and produce $H_2$. A comparison between the HEON and corresponding HEO showed the addition of nitrogen leads to a wider light absorbance, smaller bandgap and less electron-hole recombination due to changes in orbitals around the titanium and niobium elements (particularly in the conduction band) and extending the bond length for other elements. These electronic and optical changes by the addition of nitrogen lead to greater $H_2$ generation from PET photolysis, introducing HEONs as new potential catalysts for photoreforming of plastic waste.

## 2. Materials and methods
### 2.1. Reagents

Five metals with high purity, titanium 99.9%, zirconium 99%, hafnium 99.7%, niobium 99.9% and tantalum 99.9%, were purchased from Furuuchi Chemical Cooperation, Japan. Sodium hydroxide, sodium deuteroxide, hydrogen hexachloroplatinate (IV) hexahydrate, PET with a particle size of 300 μm and maleic acid were purchased from Fujifilm, Japan; Sigma-Aldrich, USA, Goodfellow, UK and Nacalai Tesque, Japan, respectively.

**2.2. Synthesis of photocatalyst**

A three-step synthesis process was performed for the production of the HEON. A Ti-Zr-Hf-Nb-Ta alloy was synthesized from the pieces of pure metals in the first step. The masses of metals were calculated so that the atomic percentages of the five elements in the HEA were similar. A mixture of metals was mixed in acetone for 30 min, dried, placed in an arc melting, melted, and homogenized with 14 times flipping and remelting to obtain an ingot with 12 g mass. After arc melting, an electric discharge machine was used to cut the ingot in the form of a 5 mm radius and 0.8 mm thick discs. The discs underwent severe plastic deformation through high-pressure torsion (turns of $N = 10$, speed of $\omega = 1$ rpm, pressure of $P = 6$ GPa and temperature of $T = 300$ K) to enhance the homogeneity and introduce lattice defects [35,36]. In the second step, the HEA was used to synthesize an HEO. The HEA was crushed and calcined for one day at 1373 K, and this process was performed two times to ensure that the entire material was completely oxidized. In the final synthesis step, the oxide was heated with a rate of 20 K/min and calcined at 1373 K under a flowing ammonia ($NH_3$) with a rate of 2.5 mL/s for 7 h to form the final HEON [15].

**2.3. Characterization of photocatalyst**

The characteristics of the HEON were examined by different methods. X-ray diffraction (XRD) using Cu Kα radiation and Raman spectroscopy analysis employing a laser with 532 nm wavelength was conducted to understand the crystal structure of the HEON. The distribution of particle size was measured using a diffraction laser particle size (DLS) analyzer. The microstructure of the catalyst was tested using a scanning electron microscope (SEM) and analysis of composition was conducted by energy-dispersive X-ray spectroscopy (EDS). The nanostructure of the material was determined by sample crushing in ethanol, dispersing it onto a carbon grid and analyzing by a transmission and scanning-transmission electron microscope (TEM and STEM) using high-resolution (HR) images, high-angle annular dark-field (HAADF) images and also EDS. X-ray photoelectron spectroscopy (XPS) with Al Kα as a radiation source was used for determining the oxidation and reduction states of elements and evaluating the valence band maximum (VBM). X-ray absorption structure (XAS) analysis using synchrotron beam light was employed to understand the local structure and the unoccupied local electronic states of elements. UV-Vis diffuse reflectance spectroscopy was carried out to determine light absorption in the 200-800 nm wavelength range. The bandgap calculation was conducted using the Kubelka-Munk analysis. Photoluminescence (PL) spectroscopy with a 325 nm wavelength laser was used to determine the recombination of electrons and holes. For the photocurrent measurement, a coating process was performed based on a previous study, where 5 mg of the HEON was fragmented in ethanol, dispersed onto a glass of fluorine-doped tin oxide (FTO) with 15×25 mm dimensions, and then calcined at 473 K for 2 h [15]. A photocurrent measurement setup was used which included a Xenon lamp (300 W with 18 kW/m$^2$) a solution of 1 M $Na_2SO_4$ for the electrolyte, a platinum counter electrode, a reference Ag/AgCl electrode and the coated FTO glass for anode electrode. The photocurrent test was performed in the potentiostatic amperometry mode with 90 s light ON and 90 s light OFF, and it was repeated continuously.

**2.4. Photoreforming experiments**

The schematic graphic of the photoreforming experiment is described in Fig. S1 of supporting information. The PET plastic with 50 mg mass and HEON with 50 mg mass were added to a test quartz tube reactor with a volume of 28 mL, followed by adding 3 mL NaOH 10 M (for alkaline hydrolysis of PET) and 250 μL $H_2PtCl_6·6H_2O$ (as a source for platinum cocatalyst). The solution was then subjected to 5 min sonication, sealed with a septum cap, and bubbled with argon gas for 15 min to remove air. The test tube was located in a cooling water bath and placed on a magnetic stirrer to maintain the disturbance. The bottom of the test tube was located 10 cm away from a Xenon lamp (with 300 W overall power, 18 kW/m$^2$ power intensity and a wavelength range of 200-3500 nm). The gas phase in the reactor was taken by a syringe with 50 μL volume and examined with a gas chromatograph equipped with a thermal conductivity detector (GC-TCD) to determine the concentration of $H_2$ produced from the reaction. To qualitatively identify the oxidation products from the photoreforming of PET plastic, the $^1$H nuclear magnetic resonance (NMR) method was performed. For preparing the NMR sample, 3 mL of 10 M NaOD in $D_2O$ was used instead of 10 M NaOH in $H_2O$, and the solution after 4 h irradiation was filtered by using a polytetrafluoroethylene (PTFE) hydrophilic filter (13 mm filter diameter and 0.45 μm pore size). The quantification of the oxidation products was performed by adding maleic acid as an internal standard before the measurement [14]. To recognize the dominant radicals and species for photoreforming, masking tests were conducted by adding 3 mL (0.02 mmol/L) of isopropanol (IPA), ethylenediaminetetraacetic acid (EDTA) and silver nitrate ($AgNO_3$) as hydroxyl radical (·OH), hole (h$^+$) and electron (e$^-$) scavengers, respectively.

**3. Results**

**3.1. Microstructure and composition characterization**

Fig. 1a and Fig.1b depict the crystal structures of HEA, HEO and HEON, respectively. HEA has only a body-centered cubic (BCC) phase, but it transforms to HEO with two monoclinic and orthorhombic phases by oxidation. The HEON has a two-phase structure containing 37% monoclinic (P21/c group, $a = 5.103$ Å, $b = 5.124$ Å, $c = 5.281$ Å; $α = γ = 90°$, $β = 99.15°$) and 63% face-centered cubic (Fm3m space group, $a = b = c = 4.296$ Å; $α = β = γ = 90°$), as shown more clearly using Rietveld refinement in Fig. 1b. Fig. 1c illustrates the particle size distribution of the HEON by DLS, ranging from 0.3 to 205.0 μm with a mean particle size of 20.6 ± 0.5 μm. Fig. 1d shows the Raman spectra of the HEON taken from four different positions. Only a distinguishable pattern is observed despite the dual-phase structure of the material, which should be due to the smaller sizes of phases compared to the spot size of Raman spectroscopy. Such structural evolutions are consistent with earlier reports [15].

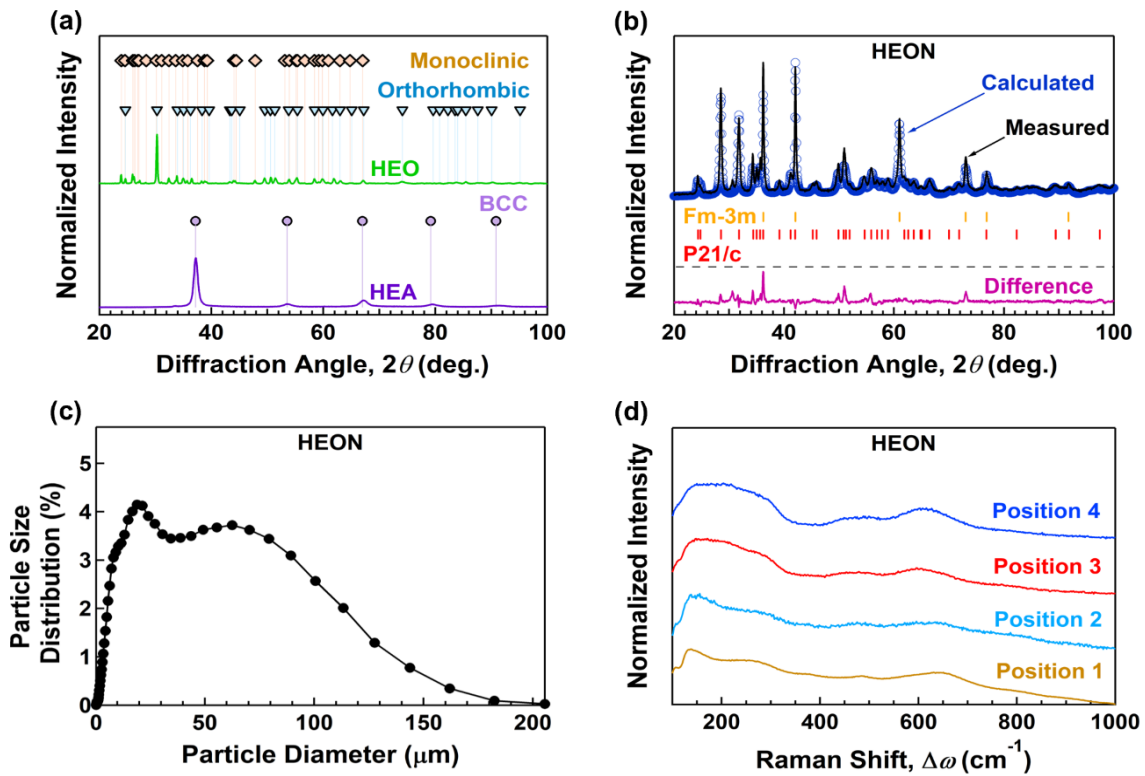

Fig. 1. Structural characterization and particle size distribution of high-entropy oxynitride photocatalyst with dual monoclinic and face-centered cubic phases. (a) XRD profile of HEA and HEO, (b) XRD profile of HEON and corresponding Rietveld analysis, (c) particle size distribution by DLS analysis of HEON, and (d) Raman spectra of HEON taken at four different positions.

An HR image by TEM is shown in Fig. 2a, confirming that the material has two distinct phases of face-centered cubic and monoclinic with nanometer sizes. It should be noted that phases were recognized using a fast Fourier transform in Fig. 2a. A higher magnification view of the microstructure shows the presence of a distorted structure, as shown in Fig. 2b and 2c. This distorted structure, which should be due to the existence of various elements with diverse atomic diameters in the lattice, is accompanied by the formation of dislocations. Although there is limited information about the effect of dislocations on catalytic activity, dislocations are sometimes considered effective defects to enhance the photocatalytic activity [37].

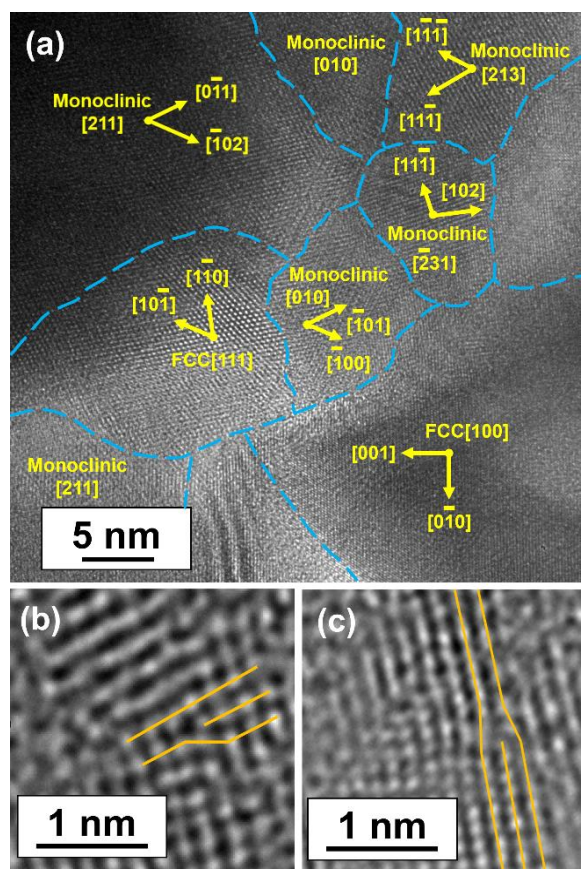

Fig. 2. Formation of two phases and significant lattice distortion in high-entropy oxynitride photocatalyst. (a) TEM high-resolution images of nanograins, and (b, c) lattice images of distorted regions containing dislocation-like defects.

The elemental distribution in the HEON, shown in Fig. 3a and Fig. 3b using SEM- and STEM-EDS mappings, indicates that there is a reasonably homogenous distribution of elements at both micrometer and nanometer levels. Despite good overall homogeneity, the distribution of titanium is not as homogeneous as the other four elements, suggesting that the presence of two phases is due to the Ti-rich and Ti-poor regions. The overall good homogeneity should be due to the effect of both thermal treatment and pre-processing by high-pressure torsion [36]. Quantitative SEM-EDS analysis suggests that the HEON contains 9.3% Ti, 6.5% Zr, 6.9% Hf, 7.4% Nb, 7.6% Ta, 41.7% O and 20.6% N (at%).

Fig. 4 indicates the oxidation state of elements in the HEON using XPS and corresponding peak deconvolution. Fig. 4 reveals the presence of $d^0$ configuration (empty d orbital) for cations ($Ti^{4+}$, $Zr^{4+}$, $Hf^{4+}$, $Nb^{5+}$ and $Ta^{5+}$) as well as the $O^{2-}$ and $N^{3-}$ anion states which is consistent with the expected characteristics of the oxynitride group [38–40]. These results demonstrate the successful full oxidation of the HEA to the HEO and the synthesis of HEON in a three-step thermomechanical oxidation process.

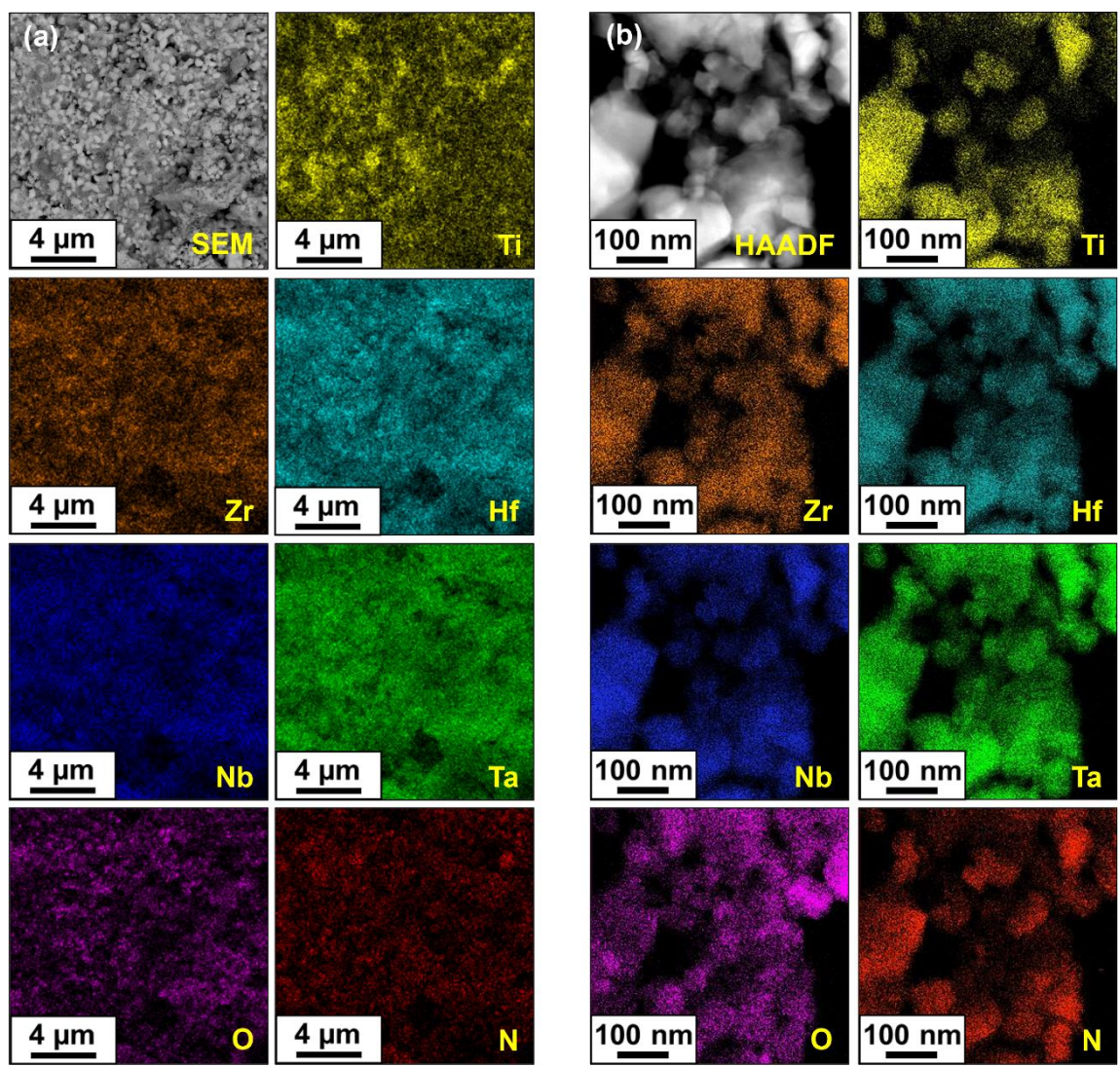

Fig. 3. Homogeneous distribution of elements in high-entropy oxynitride photocatalyst in micro/nanometer sizes. (a) SEM with EDS mappings, and (b) HAADF image taken by STEM and related EDS mappings for HEON.

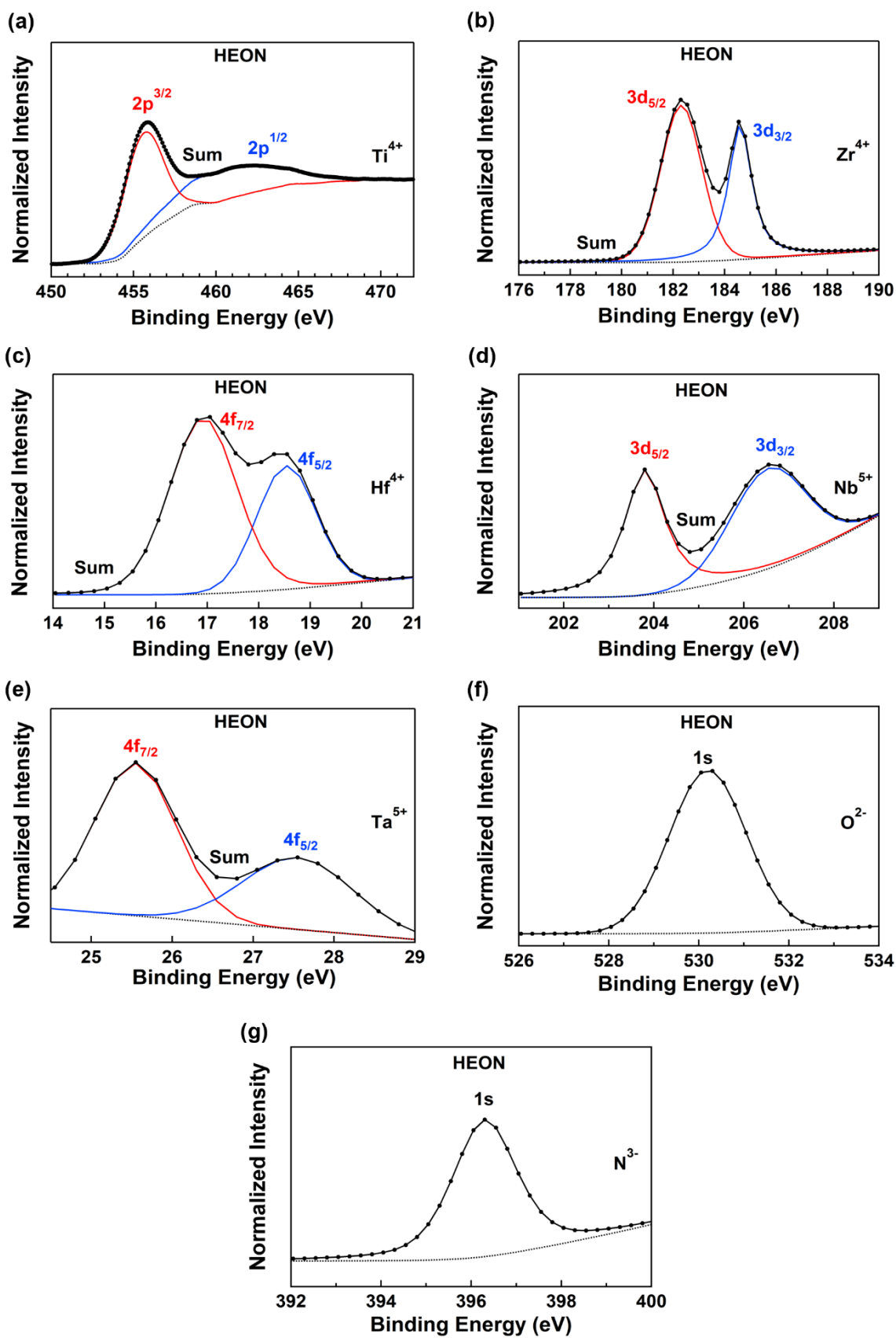

Fig. 4. Full oxidation states for metals and full reduction states for oxygen and nitrogen in high-entropy oxynitride photocatalyst. XPS spectra and relevant peak deconvolution analysis for (a) Ti$_{2p}$, (b) Zr$_{3d}$, (c) Hf$_{4f}$, (d) Nb$_{3d}$, (e) Ta$_{4f}$, (f) O$_{1s}$ and (g) N$_{1s}$ of HEON.

## 3.2. Atomic structure characterization

Fig. 5 shows the XAS spectra versus (a, c, e, g, i) energy and (b, d, f, h, j) radial distance for five metals in the HEON to clarify the atomic structure and coordination of elements. To understand the influence of nitrogen on the electronic structure, the XAS spectra of the HEO were also included in Fig. 5. It should be noted that XAS data have differences in the sudden increase in absorption at some particular X-ray photon energies for elements when using powerful and tunable X-ray beams. The absorption edges describe the excited core electron ejected into the continuum forming a photoelectron [41]. In the X-ray absorption near-edge structure (XANES), the K-edge absorption spectrum characteristic is based on photoelectron excitation from the s layer to the p layer, where p denotes the lowest unoccupied orbital of the absorbing atoms like titanium, zirconium and niobium. Similarly, the $L_3$-edge ($2p_{3/2}$) describes the excitation of electrons from the p orbital to the d orbital, which applies to the elements hafnium and tantalum. Examination of Fig. 5 indicates a few important points about the atomic structures of various elements as follows.

First, an electron transition to unoccupied final states fills the d-band configuration of zirconium, hafnium, niobium and tantalum, resulting in no pre-edge peak for these elements for both HEO and HEON in Fig. 5c, 5e, 5g and 5i. However, titanium has an empty 3d orbital ($d^0$ transition metal), which results in the appearance of the pre-edge peak at an energy of 4.97 keV, as described in Fig. 5a [42]. This energy position emphasizes the dominant 5-fold coordination of titanium in the HEO and the HEON [43]. The decrease in the intensity in the pre-edge peak without any shift in peak position indicates a higher symmetry of titanium in the HEON than in the HEO [44].

Second, considering titanium, niobium and tantalum at the rising edge in the HEON, an edge shift to the low energy ($E_0$) compared to the HEO, indicates that the valence and binding energy of the above elements are correspondingly smaller [45]. Besides, the valence of elements zirconium and hafnium are almost the same in both materials, as observed by no change in their $E_0$.

Third, the rising edge might lead to a sharp peak, commonly known as the "white line". Since white line indicates the occupation of electrons at the 5d layer, and low occupation leads to higher white line intensity and vice versa [46]. Fig. 5e and 5i indicate that Hf-d and Ta-d have lower electron density in the HEON than the HEO. Similarly, it can be concluded that Zr-p has a higher electron density in the HEON than the HEO.

Fourth, Fig. 5a and Fig. 5g also show that in the extended X-ray absorption fine structure (EXAFS) region, the structure of the titanium and niobium vicinity is completely different between the HEON and the HEO. Since these two elements are the major contributors to the conduction band minimum (CBM) [46], it is evident that the addition of nitrogen significantly modifies CBM.

Fifth, the peak position for zirconium (Fig. 5c) and tantalum (Fig. 5i) at post-edge are located on the lower energy side for the HEON compared to the HEO, indicating that the Zr-O/N and Ta-O/N bond lengths being larger than those of Zr-O and Ta-O. This contrasts with hafnium (Fig. 5e), where the peak position of the HEO is located on the lower energy side, meaning the Hf-O bond length is longer than that of Hf-O/N.

Sixth, the normalized EXAFS functions in R-space (Fourier-transformed EXAFS), plotted in Fig. 5b, 5d, 5f, 5h and 5j for titanium, zirconium, hafnium, niobium and tantalum, respectively, provide a better description about the neighboring structures of the elements. In the normalized EXAFS functions, a decrease in the amplitude of the first dominant peak at around 0.8 - 1.5 Å, corresponding to the oxygen nearest neighbor shell for Ti K-edge, Nb K-edge and Ta $L_3$-edge in the HEON can be ascribed to the increase of the oxygen vacancy around titanium, niobium and tantalum compared to the HEO [47,48]. This contrasts with the increase in the amplitude of Hf $L_3$-edge, indicating that after nitriding, the oxygen atoms in the HEON are closer to hafnium with low electronegativity, which is also consistent with the smaller bond length of Hf-O/N compared to Hf-O as mentioned earlier.

These atomic structure differences caused by the addition of nitrogen to the HEO can significantly influence the optical and electronic features of the HEON, as reported earlier for some conventional oxynitrides [43–46].

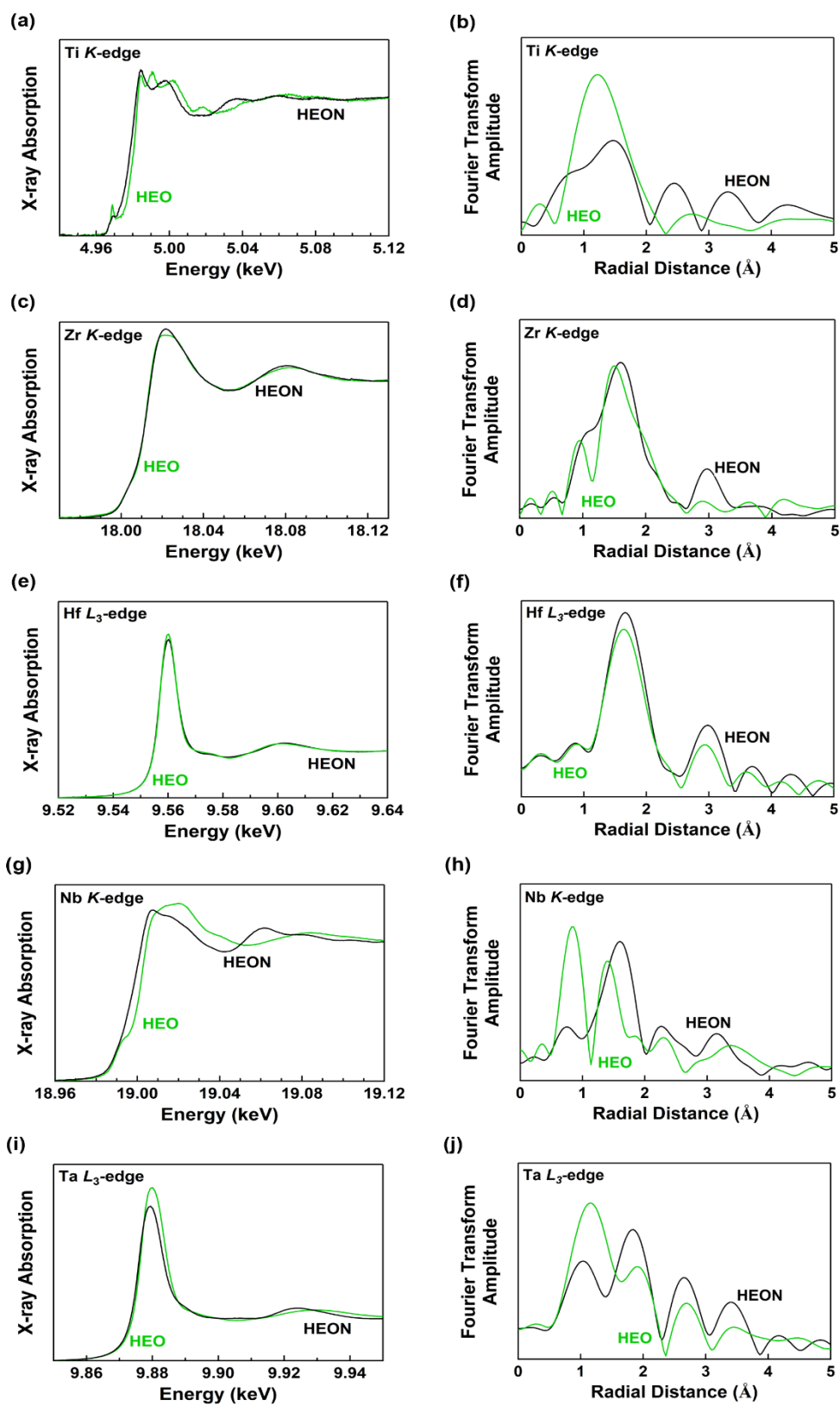

Fig. 5. The difference in the local atomic structure and vicinity of metals in high-entropy oxynitride and oxide photocatalysts. (a, c, e, g, i) XAS data and (b, d, f, h, j) Fourier-transformed extended X-ray absorption fine structure spectra for (a, b) Ti K-edge, (c, d) Zr K-edge, (e, f) Hf $L_3$-edge, (g, h) Nb K-edge, and (i, j) Ta $L_3$-edge of HEON and relevant HEO.

## 3.3. Optical property characterization

Fig. 6a illustrates the UV-Vis absorption spectrum of the HEON compared with the HEO. While the HEO exhibits light absorbance mainly in the UV region, the HEON shows high absorption intensity over a wide wavelength range from the infrared region to the UV region. The band structure of the HEON was determined by analyzing the UV-Vis spectrum using the Kubelka-Munk theory and obtaining the XPS profile at low energies. Fig. 6b illustrates the calculated graph to determine the direct bandgap and Fig. 6c illustrates the XPS spectrum to estimate VBM. The bandgap for HEON is around 1.5 eV and its VBM is approximately located at 1.4 eV vs. NHE (normal hydrogen electrode). This bandgap is significantly lower than that of the HEO which is 3.2 eV with a VBM of 1.8 eV vs. NHE. The CBM can be calculated by subtracting the bandgap from the VBM, which reaches -0.1 eV vs. NHE for the HEON and -1.4 eV vs. NHE for the HEO. Based on these calculations, Fig. 6d shows the band structure of the HEON and the HEO, indicating that the band structure of the HEON as well as of the HEO satisfies the energy requirements for water splitting (*i.e.* VBM > 1.2 eV vs. NHE and CBM < 0 eV vs. NHE).

While a low bandgap indicates easy separation of electrons and holes, the charge carrier recombination needs to be examined by methods such as photocurrent and photoluminescence measurements. Fig. 6e shows the photocurrent generation by the HEON in comparison with the HEO. Although the comparison of photocurrent intensity for these two materials is hard due to the difference in their behavior in bonding to the FTO glass, the differences in the shape of photocurrent spectra indicate some important points. A gradual increase in the photocurrent intensity over time is observed for the HEON, indicating that the photo-generated electrons can survive for a reasonable period to migrate to the catalyst surface and participate in photocurrent generation. The gradual decrease in current density after stopping irradiation also suggests the survival of charge carriers for a reasonable period. However, a spike followed by a gradual decrease in photocurrent is observed for the HEO, indicating faster recombination of charge carriers. Photoluminescence analysis shown in Fig. 6f also indicates a peak at around 575 nm for HEO; however, no peak is observed for the HEON, indicating that the electron-hole radiative recombination should be smaller for the HEON. The combination of light absorption within a wide range of wavelengths, narrow bandgap and low carrier recombination, gives this HEON a high potential as a catalyst for photoreforming [12,14].

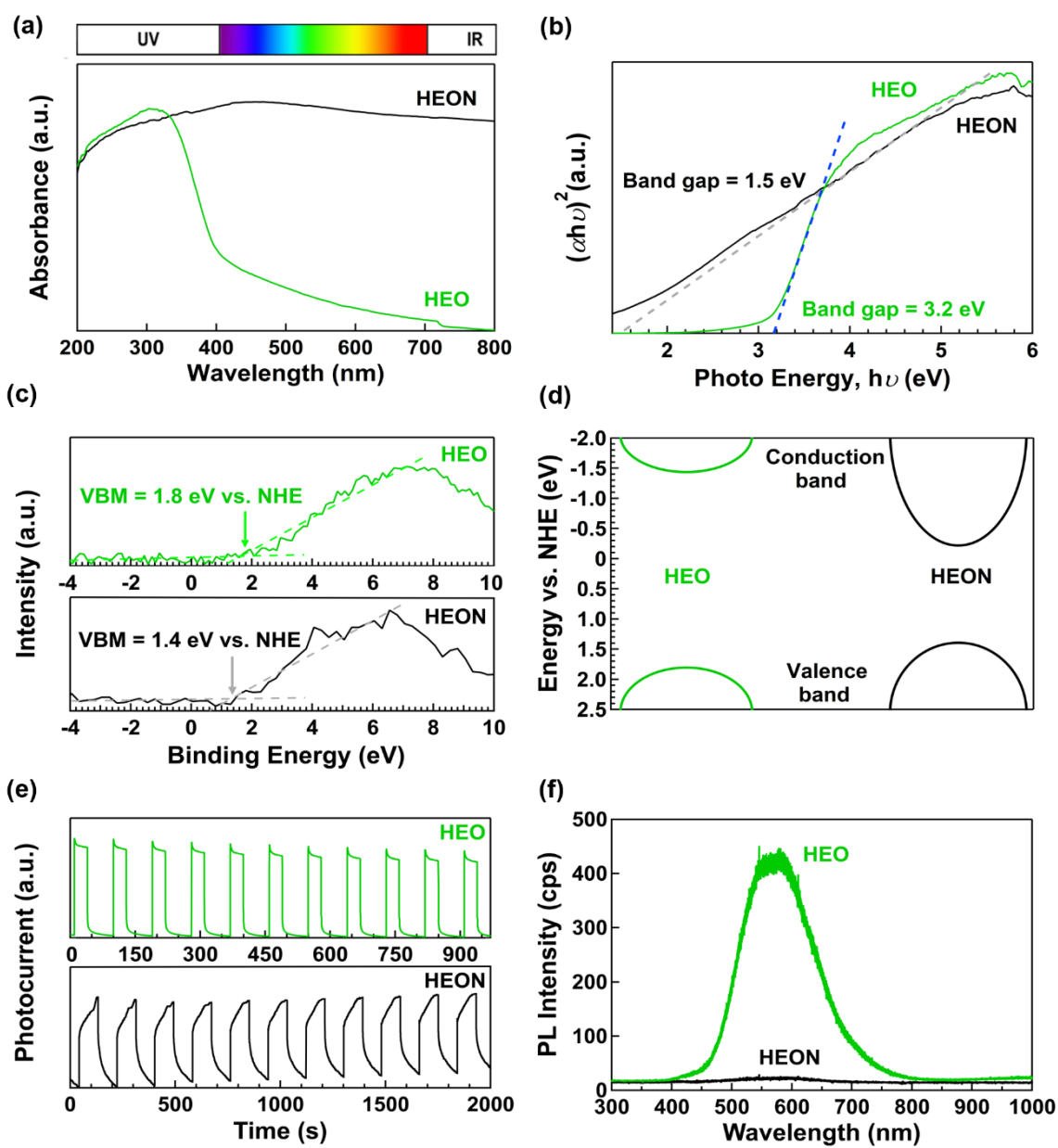

Fig. 6. Wide wavelength light absorbance, low bandgap, proper band structure, high photocurrent generation and small electron-hole recombination of high-entropy oxynitride photocatalyst. (a) UV-Vis light absorption spectroscopy, (b) Kubelka-Munk bandgap determination graph ($\alpha$: light absorbance, h: Planck's constant, $\upsilon$: frequency of light photons), (c) XPS VBM calculation, (d) electronic band structure, (e) photocurrent generation, and (f) photoluminescence spectrum achieved with a 325 nm laser for HEON in comparison with HEO.

### 3.4. Photoreforming performance

Fig. 7a shows the $H_2$ yield from the PET photoreforming using the HEON catalyst within 4 h of irradiation. For comparison, the $H_2$ production on relevant HEO was also included. Additionally, a blank experiment was also performed in the same photoreforming conditions without any catalyst addition. No $H_2$ is produced during the blank test, but it appears after the addition of HEO and HEON.

The H$_2$ production of the HEON is two times higher than that of the HEO (1.63 mmol/g versus 0.84 mmol/g), confirming that the addition of nitrogen to the HEO and the production of the HEON is effective in enhancing the catalytic efficiency for photoreforming. Fig. 7b illustrates the masking experiments for capturing reactive species and radicals in photoreforming over the HEON. The addition of all scavengers reduces the H$_2$ production rate, but the addition of IPA as a hydroxyl scavenger has less effect than the addition of EDTA as a hole scavenger and AgNO$_3$ as an electron scavenger. These masking tests, thus, suggest that electrons and holes are the most essential species for the PET photoreforming over HEON, where electrons act for H$_2$ production and holes act for direct PET oxidation as well as for the generation of hydroxyl radicals which further contribute to PET oxidation.

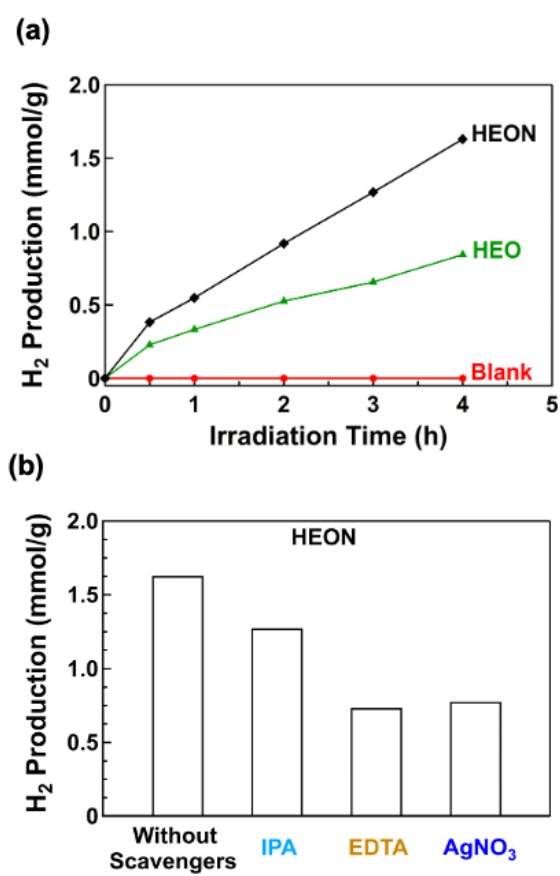

Fig. 7. Catalytic H$_2$ evolution from PET plastic degradation over high-entropy oxynitride and oxide photocatalysts, and the role of reactive species in photoreforming reaction. (a) H$_2$ production versus irradiation time for HEON compared to HEO including blank test. (b) The effect of scavengers (IPA as hydroxyl radical scavenger, EDTA as hole scavenger and AgNO$_3$ as electron scavenger) on the H$_2$ production from PET photoreforming over HEON for 4 h.

Since the half-reaction of photoreforming is PET oxidation, the oxidation products were examined via the $^1$H NMR method, as shown in Fig. 8a. The detected oxidation products are formic acid, terephthalate, ethylene glycol and acetic acid [49,50]. At high pH, PET is hydrolyzed to terephthalate and ethylene glycol [4,14] where the former product is further transformed into other organic molecules such as formic acid and acetic acid [51,52]. To determine the concentration of these products, maleic acid was added to the solution as an internal standard after the reaction, and NMR was conducted again as shown in Fig. 8b. A peak at 5.9 ppm, depicting the intensity of the internal standard, appears in Fig. 8b. Based on this internal standard, Fig. 8c illustrates the concentration of products formed from the PET degradation. It can be seen that PET degradation on the HEON leads to the formation of 0.814 mmol/g of formic acid and 0.215 mmol/g of acetic acid within 4 h irradiation. The amounts of these photodegradation products are higher compared to the HEO, which produces no acetic acid and only 0.38 mmol/g of formic acid. These findings indicate that the HEON has a high capability in plastic treatment to form valuable inorganic products.

Fig. 9 describes the stability and cycling performance of the HEON for PET degradation. The recycled HEON after four repeated cycles (total 16 h) maintains approximately 99% $H_2$ evolution efficiency compared to a fresh HEON, as shown in Fig. 9a. XRD patterns (Fig. 9b), Raman spectra (Fig. 9c) and SEM image (Fig. 9d and Fig. 9e) of the HEON before and after the photoreforming show no detectable change in the structure, demonstrating that the HEON has good chemical stability even in highly alkaline solution, possibly due to the entropy-stabilization effect by the presence of many principal elements [15].

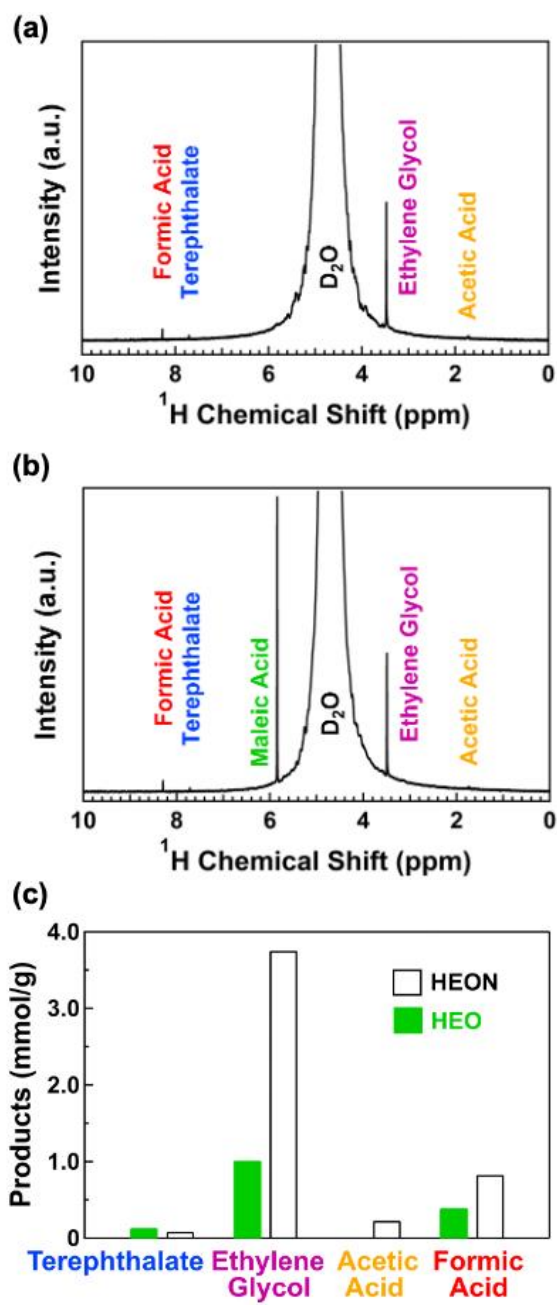

Fig. 8. Determination of oxidation products from PET plastic photoreforming using high-entropy oxynitride photocatalyst in 10 M NaOD in $D_2O$. $^1H$ NMR spectra (a) without and (b) with maleic acid addition, and (c) concentration of oxidation products from PET degradation after 4 h photoreforming using HEON compared to HEO.

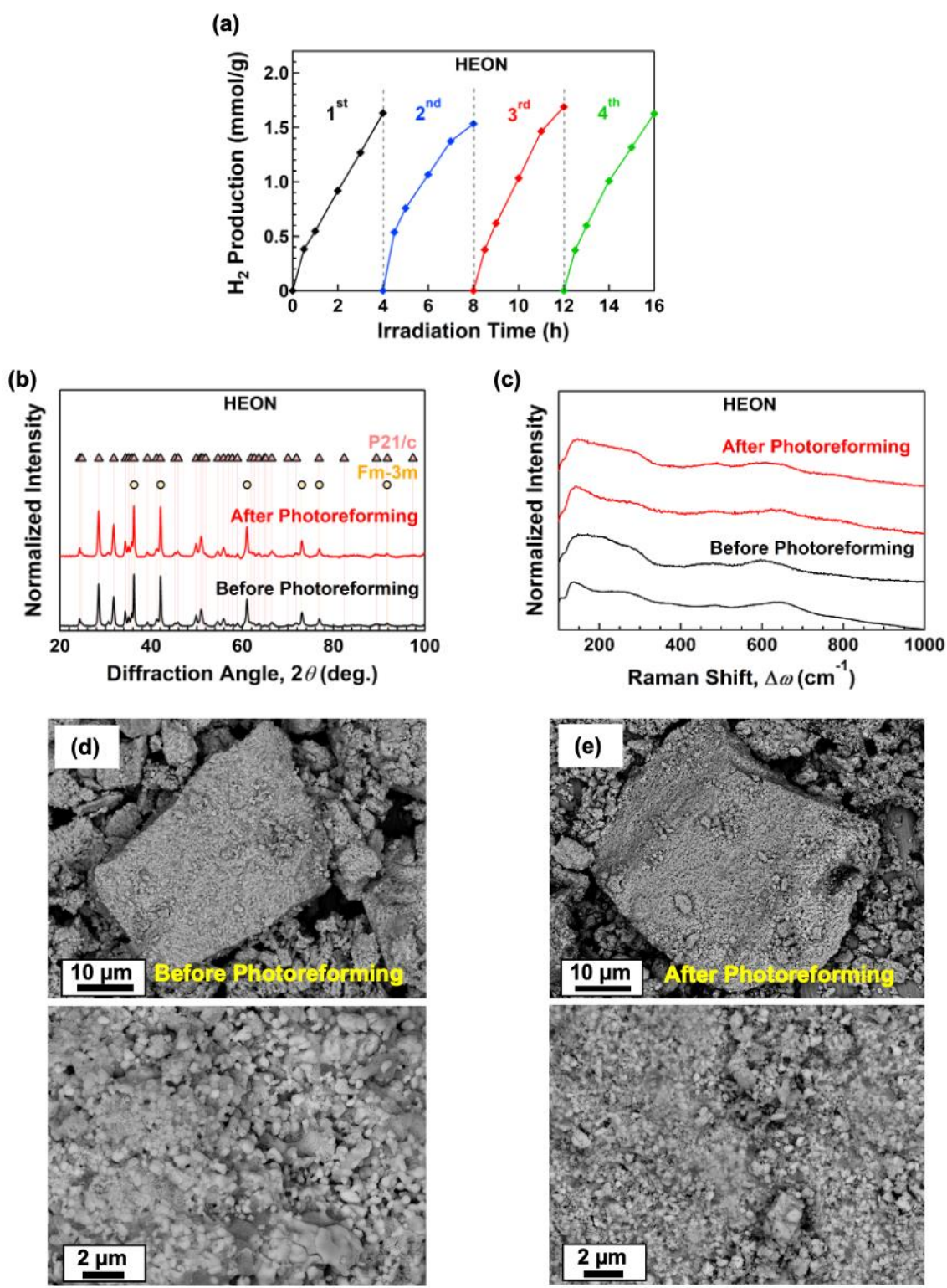

Fig. 9. Stability and reusability of high-entropy oxynitride photocatalyst for simultaneous hydrogen evolution and PET plastic photoreforming. (a) Reusability of HEON for $H_2$ production from PET photoreforming in four cycles. (b) XRD profiles and (c) Raman spectra of HEON before and after 4 h photoreforming. (d, e) SEM image of HEON (d) before and (e) after photoreforming.

## 4. Discussion

This investigation reports the first application of HEONs for the photoreforming process to degrade plastics and simultaneously produce $H_2$. Two issues are discussed in this section: (i) reasons behind the higher catalytic activity of the HEON than the HEO, and (ii) the PET degradation pathway to form valuable products.

Regarding the first issue, we need to consider the electronic and structural characteristics of the HEON and the HEO from different aspects. First, the band structure of the HEON is impacted by the hybridization of 2p orbitals from oxygen and nitrogen, with the higher N-2p orbital energy compared to O-2p orbital energy, making the bandgap of the HEON desirably narrower than the HEO [15,53]. Second, the XAS data presented in the results section show that the vicinity of titanium and niobium for the HEON significantly changes compared to the HEO. This indicates that titanium and niobium rearrange their positions in the crystal lattice while the elements zirconium, hafnium and tantalum mainly maintain the structure from the HEO. These rearrangements of atoms in the neighborhood of titanium and niobium are expected to affect the CBM of the HEON because an earlier study using first-principles calculations showed that these two elements, particularly titanium, have a significant influence on the CBM of HEOs [35]. A similar effect of nitrogen was reported for Ti-doped LaON in the literature [54]. Third, the expansion of Zr-O/N and Ta-O/N bonds and the contraction of the Hf-O/N bonds in the HEON originating from atomic disorder suggest the formation of a strained structure by the addition of nitrogen [47,55]. Such atomic-scale lattice strains can lead to the formation of defects such as vacancies by increasing the oxygen vacancy around elements (titanium, niobium and tantalum) and dislocations in larger scales (Fig. 2) which can act as activation sites for catalysis as suggested by first-principles calculations [54]. In fact, the replacement of $O_2$ sites with $N_3$ sites and the appearance of $MeN_x$ (Me: metal) bonds in oxynitrides should naturally lead to distorting the lattice and promoting vacancy formation [56]. Fourth, according to Sabatier's principle for effective catalysts, the adsorption sites on the surface must not be too weakly bonded to the reactant because activation does not occur [57]. Moreover, the adsorption sites must not be too strongly bonded because they cannot dissociate after the reaction [57]. Based on the XAS data on the reduced occupation of electrons in the valence orbital of titanium, niobium and tantalum cations by the addition of nitrogen (no change for zirconium and hafnium), it can be concluded that the overall average oxidation state of the five cations in the HEON is slightly lower than that in the HEO. This led to more appropriate binding between the active sites of HEON and the reactant in the catalytic reaction [57].

Regarding the PET photodegradation pathway and mechanism, it should be noted that the photoreforming of PET by the HEON is carried out in the absence of oxygen rather than in the presence of oxygen, unlike photodegradation and photooxidation [58]. Therefore, the photogenerated holes do not combine with oxygen to form radicals such as superoxide ($O_2^{\bullet-}$) and singlet oxygen ($^1O_2$) [59]. However, hydroxyl (•OH) radicals can be formed.

$$\text{Catalyst} + \text{Photon Energy} \rightarrow e^- + h^+ \quad (1)$$

$$h^+ + H_2O \rightarrow \bullet OH + H^+ \quad (2)$$

Since $h^+$ holes move to the catalyst surface and the lifetime of •OH is generally short, mainly •OH radicals bounded to the surface of the catalyst are formed instead of free •OH radicals in the solution [59]. Based on various studies, not only •OH radicals on the surface but also $h^+$ holes act as the oxidation agent in the photoreforming process [14,60,61]. Therefore, the PET degradation on current HEON is caused by two main agents, photogenerated holes and surface-bounded •OH radicals.

$$\text{PET} + \bullet OH \text{ or } h^+ \rightarrow \text{Organic Molecules} \quad (3)$$

Through $^1$H NMR, the products formed include terephthalate, ethylene glycol, formic acid and acetic acid, among them terephthalate and ethylene glycol are monomers of PET. It should be noted that terephthalate has a low solubility of 0.0019 g per 100 g $H_2O$ at 298 K [62] and it precipitated in the solid form during photoreforming and was filtered using a PTFE hydrophilic filter before the NMR analysis, and thus, its fraction was significantly underestimated in Fig. 8 using NMR. Additionally, terephthalate with aromatic moieties has low oxidation potential during photoreforming [4], while ethylene glycol can be oxidized further to produce smaller organic molecules such as formic acid and acetic acid by photoreforming [63]. Besides the oxidation of plastic, photo-generated electrons participate in the reduction of $H_2O$ to form $H_2$ [14].

$$2e^- + 2H_2O \rightarrow H_2 + 2OH^- \quad (4)$$

The proposed PET photoreforming pathway using the HEON can be summarized in Fig. 10. It should be noted that this pathway depends on the catalyst and photoreforming conditions, and different final products can be achieved by changing these conditions. Table 1 compares the final degradation products for photoreforming using different catalysts and testing conditions [4,14,18,52,64–68]. The amount of hydrogen production and concentration of PET oxidation products in Table 1 should be compared with care because various factors such as light source, reactant concentrations, catalyst surface area and photoreactor geometry can influence the results in different publications. It can be seen that despite the low surface area of the HEON (2.3 $m^2$/g examined by nitrogen adsorption), it shows a significant efficiency in PET photoreforming compared to other catalysts with large surface area (74.4 $m^2$/g for Pt-loaded $C_3N_4$ [18], 48.3 $m^2$/g for Pt/$TiO_2$ [65] and 22.1 $m^2$/g for porous $CN_x$ microtube [68]). In addition to the low surface area, one reason for the relatively low $H_2$ production rate using the HEON is that the CBM of the HEON is only -0.1 eV vs. NHE which generates a small overpotential for $H_2$ production. Besides, as mentioned in the previous section, one advantage of the HEONs is their high light absorbance and narrow bandgap compared to other catalysts applied for photoreforming, such as $TiO_2$ with 2.92-3.04 eV bandgap [18], $MoS_2$/CdS with 2.27 eV bandgap [52] and Pt/C$_3$N$_4$ with 2.79 eV bandgap [18]. The current results together with a recent publication about HEOs [69] confirm the high potential of high-entropy photocatalysts for photoreforming of plastic waste. Further studies should consider optimizing the characterization and surface area of HEONs to obtain the highest efficiency for the production of desired PET photoreforming products and high H$_2$ production. Moreover, the study of the photothermal conversion activity of such HEONs can be another interesting topic.

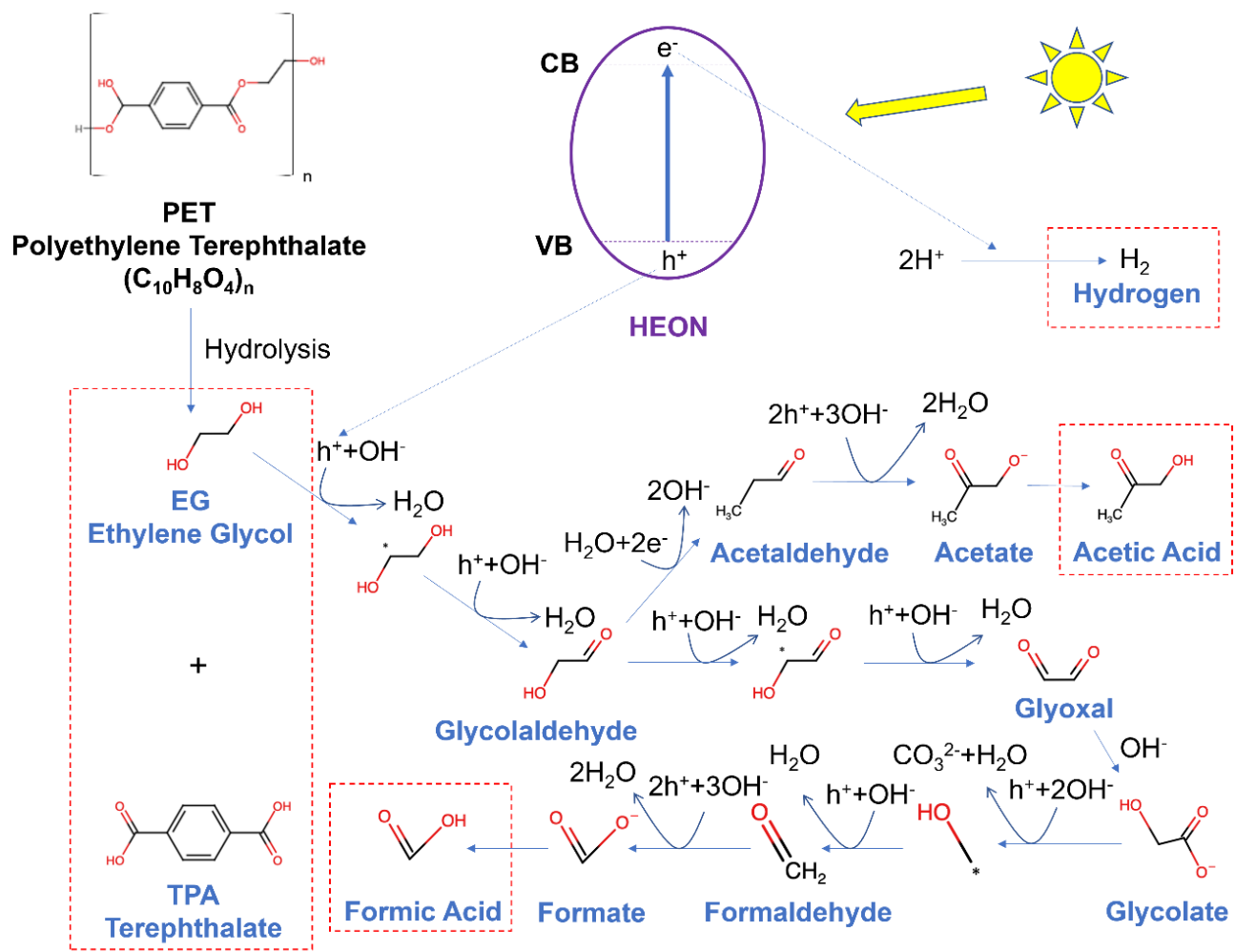

Fig. 10. The proposed pathway for PET plastic degradation by photoreforming high-entropy oxynitride photocatalyst.

Table 1. Summary of some studies for photoreforming products using different catalysts and processing conditions.

| Photocatalyst (Concentration) | Plastic (Concentration) | Conditions | Hydrogen production | Oxidative production | Reference |
|---|---|---|---|---|---|
| HEON (50 mg/3.25 mL) | PET (50 mg/3.25 mL) | - Solution: 3 mL NaOH 10 M, 250 µL $H_2PtCl_6 \cdot 6H_2O$<br>- Light: Xenon lamp (18 kW/m$^2$)<br>- Reaction time: 3 h | 0.543 mmol/h.g (0.236 mmol/h.m$^2$) | Formic acid: 0.271 mmol/h.g<br>Acetic acid: 0.072 mmol/h.g | This study |
| $CN_x/Ni_2P$ (1.6 mg/mL) | PET (25 mg/mL) | - Solution: 2 mL KOH 1 M<br>- Light source: Simulated solar light (AM 1.5G, 100 mW.cm$^{-2}$)<br>- Reaction time: 20 h | 0.026 mmol/h.g | Acetic acid 0.003 mmol/h.g<br>Formic acid 0.003 mmol/h.g<br>Glyoxal 0.145 mmol/h.g<br>Glycolate | [4] |
| $TiO_2$ anatase, brookite, rutile (50 mg/3.25 mL) | PET (50 mg/3.25 mL) | - Solution: 3 mL NaOH 10 M, 250 µL $H_2PtCl_6 \cdot 6H_2O$<br>- Light source: Xenon lamp (18 kW/m$^2$)<br>- Reaction time: 3 h | 1.477 mmol/h.g (anatase), 2.27 mmol/h.g (brookite), 1.723 mmol/h.g (rutile) | Acetic acid: 10.2 mmol/h.g (anatase), 3.3 mmol/h.g (brookite), 3.31 mmol/h.g (rutile) | [14] |
| Pt-loaded g-$C_3N_4$ (100 mg/50 mL of $H_2O$) | 50 mL of PET (2 g/100 mL) | - Solution: 100 mL NaOH 5 M, 72 h stirred at 343 K<br>- Light source: Solar light simulator (100 mW.cm$^{-2}$)<br>- Reaction time: 192 h | 7.33 mmol/h.g (0.103 mmol/h.m$^2$) | Formic acid<br>Glyoxal<br>Glycolate<br>Acetic acid | [18] |
| $MoS_2/CdS$ (100 mg/60 mL) | PET (1.5 g/60 mL) | - Solution: 60 mL KOH 10 M, 48 h stirred at 313 K<br>- Light source: Xenon lamp (300 W) with an AM 1.5 solar simulator<br>- Reaction time: 25 h | 3.9 mmol/h.g | Formic acid: 0.143 mmol/h.g<br>Acetic acid: 0.19 mmol/h.g$_{cat.}$<br>Glycolate. | [52] |
| $Pt/TiO_2$ (10 mg/2 mL) | PET bottle (125 mg/50 mL) | - Solution: 52 mL NaOH 1 M, 24 h stirred at 423 K<br>- Light source: Xenon lamp (300 W, $\lambda$ = 320-780 nm)<br>- Reaction time: 3 h | 0.219 mmol/h.g | Acetic acid: 0.143 mmol/h.g | [64] |
| Mesoporous $ZnIn_2S_4$ (2.5 mg/mL) | PET (25 mg/mL) | - Solution: 5 mL KOH 1 M, 24 h stirred at 313 K<br>- Light source: Solar light simulator (150 mW.cm$^{-2}$)<br>- Reaction time: 12 h | 0.13 mmol/h.g$_t$ | Lactide<br>Lactic acid | [65] |
| $Ni_2P/ZnIn_2S_4$ (5 mg/10 mL) | PET (5 mg/10 mL) | - Solution: 10 mL KOH 5 M, 48 h stirred at 313 K<br>- Light source: White LED light ($\lambda$ > 420 nm, 4 × 25 W)<br>- Reaction time: 5 h | 0.686 mmol/h.g | Glycolate<br>Formic acid<br>Acetic acid | [66] |
| $CdS/CdO_x$ (25 mg/mL) | PET (25 mg/mL) | - Solution: 2 mL NaOH 10 M, 24 h stirred at 313 K<br>- Light source: Solar light simulator (100 mW.cm$^{-2}$)<br>- Reaction time: 4 h | 3.42 mmol/h.g | Formic acid<br>Glycolate<br>Ethanol<br>Acetic acid<br>Lactate | [67] |
| $CN_x$ porous microtube (20 mg/25 mL) | PET (12.5 mg/250 mL) | - Solution: 250 mL NaOH 2 M, 24 h stirred at 313 K<br>- Light source: Xenon lamp (300 W)<br>- Reaction time: 4 h | 9.39 µmol/h.g (0.0004 mmol/h.m$^2$) | Glyoxal<br>Glycolate<br>Acetic acid<br>Formic acid | [68] |

## 5. Conclusion

In this investigation, an entropy-stabilized oxynitride catalyst was developed by adding nitrogen to a Ti-Zr-Hf-Nb-Ta-based high-entropy oxide and applied for the photoreforming of plastics for the first time. The addition of nitrogen led to unique electronic modifications such as changing electronic structure in the vicinity of titanium and niobium, the expansion of Zr-O/N and Ta-O/N bond length, the contraction of Hf-O/N bond length, the decrease in valence band electrons of titanium, niobium and tantalum cations, and the hybridization of 2p orbitals of oxygen and nitrogen. These electronic structure modifications, examined by X-ray absorption near edge structure (XANES) and extended X-ray absorption fine structure (EXAFS), led to improved optical properties such as narrowing bandgap and diminishing electron-hole recombination. The plastic photoreforming to formic acid and acetic acid with simultaneous $H_2$ production was successful over the oxynitride with a higher activity compared to the relevant high-entropy oxide. These results offer an effective prospect for developing highly efficient and stable catalysts for treating plastic waste by using the concept of high-entropy oxynitrides which have both low bandgap and high stability.

**CRediT authorship contribution statement**

All authors contributed to conceptualization, investigation, methodology, validation, and writing - review & editing.

**Declaration of competing interest**

The authors declare no conflict of interest.

**Acknowledgments**

The author H.T.N.H. thanks Yoshida Scholarship Foundation (YSF) for a Ph.D. scholarship. This study is supported partly by Mitsui Chemicals, Inc., Japan, partly through a Grant-in-Aid from the Japan Society for the Promotion of Science (JP22K18737), and partly by the ASPIRE project of the Japan Science and Technology Agency (JST) (JPMJAP2332).

# Supplementary Information

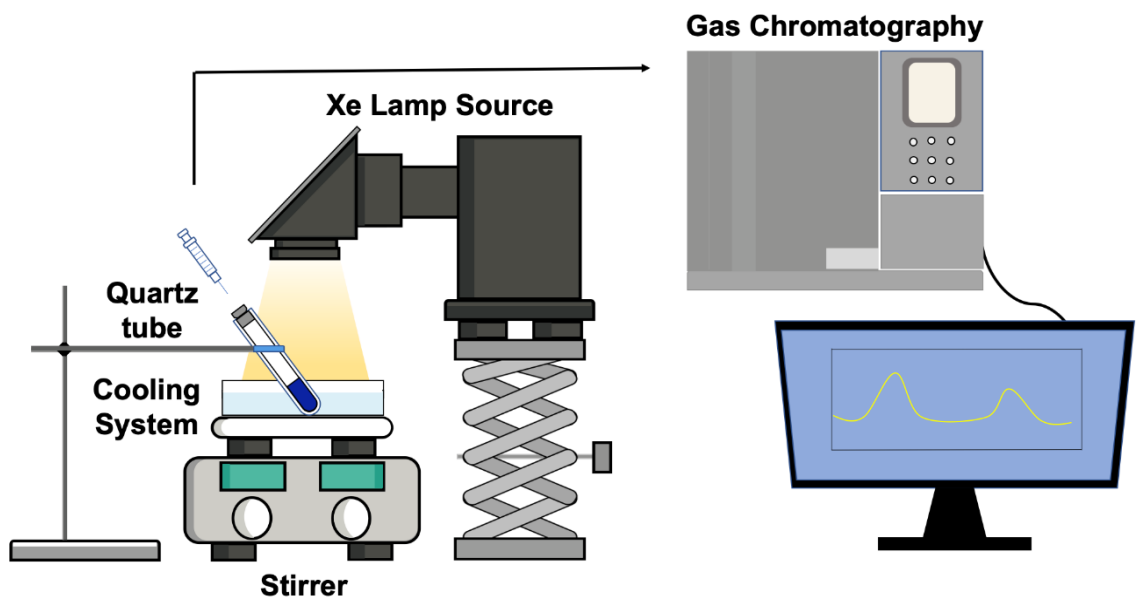

Figure S1. Schematics of photoreforming experiment.